\documentclass[10pt,aps,prl,nofootinbib,superscriptaddress,twocolumn,preprintnumbers,balancelastpage]{revtex4-1}
\usepackage{amsmath,amssymb,amsthm,amsopn}
\usepackage{epsfig}
\usepackage{psfrag}
\usepackage[colorlinks=true,allcolors=blue!65]{hyperref}
\usepackage{multirow}
\usepackage{empheq}
\usepackage{slashed}
\usepackage{graphicx}
\usepackage{url}
\usepackage{subcaption}
\usepackage{textcomp}
\usepackage{bm}
\usepackage{dcolumn}
\usepackage{color,xcolor}
\usepackage{ulem}
\usepackage{cancel}
\usepackage{braket}
\usepackage[title]{appendix}
\usepackage{lipsum}
\usepackage{mathptmx}
\DeclareMathAlphabet{\mathcal}{OMS}{cmsy}{m}{n}

\def\comment#1{}

\def\beq{\begin{equation}}
	\def\eeq{\end{equation}}
\def\bea{\begin{eqnarray}}
	\def\eea{\end{eqnarray}}

\begin{document}
	
	\title{Probing Virtual Axion-Like Particles by Precision Phase Measurements}

	\author{Moslem Zarei}
	\email[]{m.zarei@cc.iut.ac.ir}
	\affiliation{Department of Physics, Isfahan University of Technology, Isfahan 84156-83111, Iran}
	\affiliation{Dipartimento di Fisica e Astronomia \textquotedblleft G. Galilei\textquotedblright, Universit\`{a} degli Studi di Padova, via Marzolo 8, I-35131, Padova, Italy}
	\affiliation{ICRANet-Isfahan, Isfahan University of Technology, 84156-83111, Iran}
	
	\author{ Soroush Shakeri}
	\email[]{s.shakeri@ipm.ir}
	\affiliation{Department of Physics, Isfahan University of Technology, Isfahan 84156-83111, Iran}
	
	\affiliation{ICRANet-Isfahan, Isfahan University of Technology, 84156-83111, Iran}
	
	\affiliation{School of Astronomy, Institute for Research in Fundamental Sciences (IPM), P. O. Box 19395-5531, Tehran, Iran}
	
		\author{Mohammad Sharifian}
	\email[]{mohammadsharifian@ph.iut.ac.ir}
	\affiliation{Department of Physics, Isfahan University of Technology, Isfahan 84156-83111, Iran}
	\affiliation{ICRANet-Isfahan, Isfahan University of Technology, 84156-83111, Iran}
	
	\author{Mehdi Abdi}
	\email[]{mehabdi@cc.iut.ac.ir}
	\affiliation{Department of Physics, Isfahan University of Technology, Isfahan 84156-83111, Iran}
	
	\author{David J. E. Marsh}
	\email[]{david.marsh@uni-goettingen.de}
	\affiliation{King's College London, Strand, London, WC2R 2LS, United Kingdom}
	
	\author{Sabino Matarrese}
	\email[]{sabino.matarrese@pd.infn.it}
	\affiliation{Dipartimento di Fisica e Astronomia \textquotedblleft G. Galilei\textquotedblright, Universit\`{a} degli Studi di Padova, via Marzolo 8, I-35131, Padova, Italy}
	\affiliation{INFN Sezione di Padova, via Marzolo 8, I-35131, Padova, Italy}
	\affiliation{INAF-Osservatorio Astronomico di Padova, Vicolo dell'Osservatorio 5, I-35122 Padova, Italy}
	\affiliation{Gran Sasso Science Institute, viale F. Crispi 7, I-67100, L'Aquila, Italy}

	\date{\today}
	
	\begin{abstract}
		We propose an experiment  for detecting Axion-Like Particles (ALPs) based on the axion-photon interaction in the presence of a non-uniform magnetic field. The impact of virtual ALPs on the polarization of the photons inside a cavity is studied and a detection scheme is proposed.
		We find that the cavity normal modes are dispersed differently owing to their coupling to the ALPs in the presence of a background magnetic field. This birefringence, in turn, can be observed as a phase difference between the cavity polarization modes.
			The signal is considerably enhanced for a squeezed  light source. We argue that the amplified signal allows for exclusion of a range of axion mass $6\times10^{-4}\;\text{eV} \lesssim m_{a} \lesssim 6\times10^{-3}$~eV even at very small axion-photon coupling constant with the potential to reach sensitivity to the QCD axion.
		Our scheme allows for the exclusion of a range of axion masses that has not yet been covered by other experimental techniques.
		
	\end{abstract}
	
	\maketitle
	
	Axions are pseudo-scalar particles beyond the Standard Model (SM) of particle physics, originally introduced by the Peccei-Quinn mechanism in order to solve the  CP problem of QCD ~\cite{Peccei1977,Weinberg1978,Wilczek1987,Chen2007}. Moreover, Axion Like Particles (ALPs) are generalizations of the QCD axions which arise from  string theory due to the compactification of extra dimensions \cite{Svrcek:2006yi,Witten:1984dg}.
	The main difference between these two types of axions is that for ALPs,  the mass and coupling constant are  independent of each other. This property allows a much wider parameter space and hence a rather rich phenomenology for ALPs.
	Both axions and ALPs are prominent candidates for Dark Matter (DM)~\cite{Abbott:1982af,Dine:1982ah,Preskill:1982cy} and can have a multitude 
	of different couplings to the SM.
	While the allowed mass region of DM axion is limited to $10^{-24}\mathrm{eV} \lesssim   m_{a}   \lesssim   10^{-1}\mathrm{eV}$ \cite{Hlozek:2017zzf,Archidiacono:2013cha}, an ALP which is not DM can take any value of the mass, depending on the coupling strength.
	
	Extensive experimental efforts have been conducted to search for axions and ALPs in recent years \cite{Marsh:2015xka,Graham:2015ouw,Graham:2013gfa,Arias:2012az,Caldwell2017,Millar:2016cjp,MADMAX2017,Kahn:2016aff,Ouellet:2018beu,Du:2018uak,Anastassopoulos:2017ftl,Barbieri:2016vwg,Marsh:2018dlj,Lawson:2019brd,Sch_tte_Engel_2021,PhysRevLett.123.121601}. Many experiments exploit the axion coupling to two photons, which causes two commonly exploited effects; i) axion-photon mixing induced by a constant magnetic field; ii) vacuum birefringence due to the photon propagation through a distribution of axion background.
	The axion-photon conversion under magnetic field known as inverse Primakov process, has been used in order to design high-precision  haloscope cavities~\cite{Sikivie1983,Sikivie1985,Asztalos2001,Hoskins2011,McAllister2016,Anastassopoulos:2017ftl,2018arXiv180607141G,Arias2016,Asztalos2010}. Haloscopes are used to search for axion DM, and the signal is a function of the local DM density. Helioscopes, on the other hand, are  designed to search for axions produced in the sun by converting them to X-rays in a laboratory magnetic field~\cite{Anastassopoulos:2017ftl,Irastorza:2011gs}. As well as these also light-shining through a wall (LSW) experiments like ALPS~\cite{EHRET2010149,B_hre_2013} try to produce and detect ALPs via the interaction of them with photons.
	
	Thanks to the recent technological advances in laser facilities and high precision optical setups, several projects, such as BRFT  \cite{Cameron1993}, PVLAS \cite{Zavattini2008}, BMV \cite{Battesti2008}, OSQAR \cite{Pugnat_2008}, and Q$\&$A \cite{Chen2007}, have been proposed in order to find indirect evidence of axions. These kinds of searches are mainly based on looking for unexpected changes in the amplitude, phase, or polarization of propagating probe photons \cite{Raffelt:1987im,Maiani:1986md}.
	Recently, a new class of experiments has emerged, based on laser interferometry ~\cite{Melissinos:2008vn,Michimura2013,DeRocco:2018jwe,Obata2018,2018arXiv180901656L,Nagano:2019rbw,Grote:2019uvn}, where an optical cavity is used to measure the induced phase difference between two circularly-polarized modes in a laser beam due to the interaction with a background of ALP dark matter.
	Furthermore, ALPs may mediate short-range spin dependent forces between spin polarized sources such as electron and nucleon \cite{Moody:1984ba,Dobrescu:2006au,arvanitaki2014:dfa,geraci2017progress,crescini:2021,crescini:2017,crescini:2016}. Searching for ALPs in this manner is independent of the ALP DM density~\cite{Tullney:2013wqa,Arvanitaki:2014dfa}.
	
	Photon-photon scattering intermediated by a virtual axion can also provide a strategy for the detection of ALPs \cite{1999NuPhS..72..201B,Moulin:1996vv,Evans:2018qwy,Bogorad:2019pbu}. It turns out that the interaction of a photon with an external magnetic field induced by an ALP contributes to vacuum birefringence \cite{DellaValle:2015xxa}. Vacuum birefringence is a manifestation of nonlinear photon-photon scattering and causes a phase difference between polarization modes of  photons propagating through the magnetic field. In other words, the pseudo-scalar nature of the axion-photon interaction generates polarization asymmetry, and can produce a net circular polarization from an initially linearly
	polarized radiation. The s-channel of photon-photon scattering can provide mass information about ALPs with a precision of the monochromaticity of the light. Providing sufficient energy from incoming photons leads to a resonance pole where real ALPs may be created.

	In this Letter, we propose a novel experiment based on the forward scattering of photons via virtual ALP exchange from an inhomogeneous magnetic field inside a cavity [Fig. (\ref{fig:scheme})]. Such \textit{off-shell} ALPs-induced scatterings can be observed  as a phase shift in the output field of the cavity modes.
	Unlike the previous works where the scattering occurs in a constant magnetic field \cite{DellaValle:2015xxa,Cameron1993,Chen2007} or a time-varying magnetic field~\cite{sharifian2021probing}, our setup offers a momentum exchange between the photons and a non-uniform background magnetic field which gives rise to resonant scattering processes in the presence of massive ALPs.
	The resonance enhances the birefringence and thus the phase deviation signal. Since it uses off-shell ALPs, our detection scheme is not sensitive to the local DM density of ALPs.
	By creating a harmonic profile for the magnetic field inside the cavity, we show that one is able to cover a mass region by changing the incident photon energy and thus scanning the resonance.
	
	\begin{figure}[t]
		\centering
		\includegraphics[width=\columnwidth]{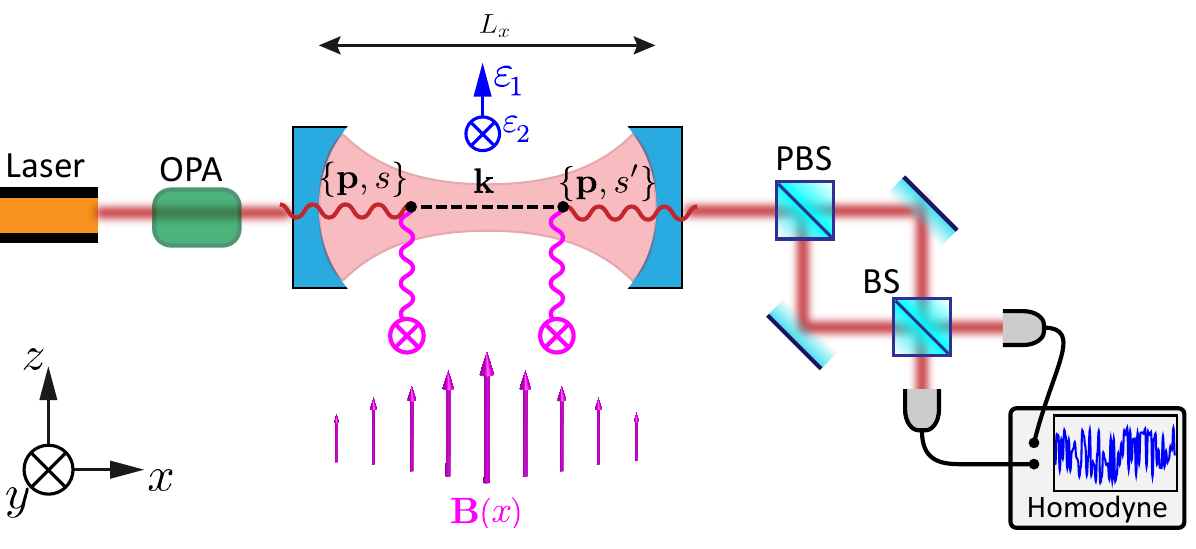}
		\caption{The scheme proposed in this paper: An optical cavity of length $L_x$ is pumped by coherent or squeezed light. A background magnetic field is applied perpendicular to the cavity axis, mediating the laser interaction with ALPs. The axion-induced dispersion introduces a phase deviation which is measured through a homodyne detection. One of the photon polarizations $\bm\varepsilon^1$ is in the direction of the magnetic field and the other one $\bm\varepsilon^2$ is perpendicular to it.}
		\label{fig:scheme}
	\end{figure}
	
	\textit{\it The framework.---}%
	Pseudoscalar ALPs couple to electromagnetic  fields 
	through the following interaction Lagrangian
	~\cite{Peccei1977,Weinberg1978,Wilczek1987}
	\begin{eqnarray}\label{int}
		\mathcal{L}_{a \gamma\gamma}=\frac{1}{4}g_{a \gamma\gamma}\,a\, F_{\mu \nu} \widetilde{F}^{\mu \nu},
	\end{eqnarray}
	where $g_{a \gamma\gamma}$ is the axion-photon coupling constant, $a$ is the pseudoscalar field describing ALPs and   $F_{\mu \nu} $(\ $\tilde F^{\mu\nu} \equiv \frac{1}{2}\epsilon^{\mu\nu\rho\sigma}F_{\rho\sigma}$) is the electromagnetic (dual-) field strength tensor. The interaction of photons with the external magnetic field can be explored by expanding the field strength tensor around the background field as $F_{\mu \nu}=\overline{F}_{\mu\nu}+f_{\mu \nu}$ where $f_{\mu\nu}=\partial_{\mu}A_{\nu}-\partial_{\nu}A_{\mu}$ behaving as a quantum field  on top of the static classical background $\overline{F}_{\mu\nu}$. Therefore, the coupling of ALPs to photons in the presence of an external background field is represented as
	\beq\label{qB}
	\mathcal{L}_{a\gamma B}=\frac{1}{2}g_{a\gamma \gamma}\,a \,\epsilon^{\mu\nu\alpha\beta} \overline{F}_{\mu\nu}\partial_{\alpha}A_{\beta}~.
	\eeq
	Interaction of a photon with an external magnetic field mediated by an ALP  can be described by the second-order interaction Hamiltonian $H_{\gamma  B}$. This Hamiltonian is effectively defined by the second-order S-matrix element $S^{(2)}(\gamma_s\rightarrow\gamma_{s'})=-i\int\! dt \,H_{\gamma  B}$ \cite{Kosowsky:1994cy}. Fig.~\ref{fig:scheme} shows the scheme of the experiment as well as the Feynman diagram of this process. Applying the forward scattering condition in which  
	the photon momenta remain unchanged (${\bf p}={\bf p'}$) while their polarization can change ($s\neq s'$), the interaction Hamiltonian is obtained as (see Supplementary Material for details)
		\begin{eqnarray}\label{hamiltonian12}			H_{\gamma B}&=&\frac{g^2_{a\gamma\gamma}}{2V}\int\frac{d^3\textbf{k}}{(2\pi)^3} d^3\textbf{x}' d^3\textbf{x}\sum_{s,s',\mathbf{p}}
			\frac{\omega_{\mathbf{p}} a_{s'}^{\dag}({\bf p}) a_s({\bf p})}{k^2-m_{a}^2+i\omega_\textbf{p}\Gamma_{a}}
			\nonumber \\ 
			&\times&
			[\bm{\varepsilon}^s\cdot\mathbf{B}(x)]
			[\bm{\varepsilon}^{s'}\!\cdot\mathbf{B}(x')]\sin(p_{x} x'+\frac{p_x L_x}{2})\sin(p_{x} x+\frac{p_x L_x}{2})\nonumber\\
			&\times&e^{-i\textbf{k}\cdot (\textbf{x}'-\textbf{x}) }\left(
			e^{-i \textbf{p}_{\bot}\cdot   \textbf{x}_\bot}
			e^{i \textbf{p}_\bot\cdot   \textbf{x}'_\bot}+
			e^{-i \textbf{p}_{\bot}\cdot   \textbf{x}'_\bot}
			e^{i \textbf{p}_\bot\cdot   \textbf{x}_\bot}\right),
	\end{eqnarray}
	where $B^k(x)=-\epsilon^{k i j}\bar{F}_{ij}(x)/2$ is the external magnetic field vector, $\varepsilon^{s}$ are the photon polarization vectors, $p=(\omega_{\mathbf{p}},\mathbf{p})$ is the  four-momentum of the pumped photons into the cavity, $k=(k^0,\mathbf{k})$ is the  four-momentum of ALPs, $V$ is the cavity mode volume, and $L_x$ is the cavity length aligned with the $x$-axis. The decay rate, $\Gamma_{a}$,  appeared in Eq.~(\ref{hamiltonian12}), due to the unstable nature of ALPs. Here, we only consider ALP masses below $2m_{e}$, where the decays into photons are only allowed~\cite{Bauer:2018uxu}. In the presence of the background magnetic field, the axion decay rate can be decomposed into $\Gamma_a=\Gamma_0(a\rightarrow \gamma \gamma)+\Gamma_B(a\rightarrow \gamma)$ where $\Gamma_0=g_{a\gamma\gamma}^2m_{a}^3/64\pi$ \cite{Raffelt:1990yz} is the axion to two photon decay rate, and $\Gamma_B$ is the axion photon conversion rate. In the high magnetic fields considered in this Letter $\Gamma_{\!B}\gg\Gamma_0$ and therefore in the following we set $\Gamma_{\!a}\approx\Gamma_{\!B}$. While our expected signal scales as $g^2_{a\gamma\gamma}/m_{a}^{2}$ for ALP masses sufficiently far from the resonance,  close to the resonant point the signal will be enhanced as $\mathcal{A}g^2_{a\gamma\gamma}$ with $\mathcal{A}$ denoting an amplification factor due to the resonance effect.
	
Note that  ALPs  with masses larger than the typical energy scale of scattered photons ($m_{a}\gg \omega_\textbf{p}$)
can be integrated out giving rise to an effective local interaction between photons \cite{Evans:2018qwy,Bogorad:2019pbu}. The experimental design in \cite{Bogorad:2019pbu} is based on 4-photon interaction in order to detect a frequency shift between photons inside a radiofrequency cavity. It is shown that the photon signal of \textit{off-shell} ALPs
is proportional to $\mathcal{N}_{a}^{2}g^4_{a\gamma\gamma}$  where $\mathcal{N}_{a}$ is the number of ALPs \cite{Bogorad:2019pbu}. Moreover, the effect of \textit{on-shell} ALPs  in Light-Shining through a Wall (LSW) experiment is proportional to $\mathcal{N}_{a}g^4_{a\gamma\gamma}$ \cite{Dobrich:2013mja,Redondo:2010dp,Redondo:2010dp}. However, our effect goes like $g^2_{a\gamma\gamma}$ rather than $g^4_{a\gamma\gamma}$ due to the fact that our detection scheme is based on measuring the phase difference. Specifically, this
	phase difference relies not on the cross section (which is proportional to square of $g^4_{a\gamma\gamma}$), but on the scattering amplitude
	$g^2_{a\gamma\gamma}$~\cite{Kotkin:1997,Hoseinpour:2020}.

	It is essential to distinguish the ALP effect from  competing processes. It is well-known that in the SM the QED  vacuum is made up of virtual pairs and  effectively behaves as a birefringent medium in the presence of electromagnetic  fields. The presence of the virtual electron-positron pairs induces nonlinear effects to Maxwell's equations, caused by  photon-photon scattering. This kind of non-linearity
	for the low-energy photons,  is encoded in the effective Euler-Heisenberg (EH) Lagrangian \cite{Euler1936, Heisenberg1936, Dicus1998,Dunne:2012hp}.  The one-loop EH Lagrangian in the presence of the background field $\bar{F}_{\mu \nu}$ is given ~\cite{Shakeri:2017knk,Shakeri:2017iph}. 
	\begin{eqnarray}\label{e23}
		\mathcal{L}_{EH}=\frac{\alpha^{2}}{90m^{4}_{e}}[5f_{\mu\nu}f^{\mu\nu}\bar{F}_{\lambda\rho}\bar{F}^{\lambda\rho}+10\bar{F}_{\mu\nu}f^{\mu\nu}f_{\lambda\rho}\bar{F}^{\lambda\rho}  
		\nonumber \\ -14f
		_{\mu\nu}\bar{F}^{\nu\lambda}f_{\lambda\rho}\bar{F}^{\rho\mu}-28f_{\mu\nu}f^{\nu\lambda}\bar{F}_{\lambda\rho}\bar{F}^{\rho\mu}]~.
	\end{eqnarray}
	
	Imposing the forward scattering condition, the interaction Hamiltonian is obtained as
	\begin{eqnarray}
		\label{e83}
		H^{EH}_{\gamma B}=-\frac{\alpha^{2}}{15m^{4}_{e}L_x}\hspace{-1mm}\int \hspace{-1mm}dx\!\sum_{\mathbf{p},s,s'}\hspace{-1mm}\omega_{\mathbf{p}} a_{s'}^{\dag}({\bf p}) a_s({\bf p})
		[\bm{\varepsilon}^s\cdot\mathbf{B}(x)][\bm{\varepsilon}^{s'}\cdot\mathbf{B}(x)]~. \nonumber\\
	\end{eqnarray}
	where $L_x$ is the cavity length. Comparing the contribution of ALP to  that coming from EH interaction  for low energy photons ($  \omega_\textbf{p} \ll m_{e}$ $and$ $m_{a} $), it can be shown that in a wide range of the parameter space the ALP effect is the dominant process \cite{Bogorad:2019pbu,Evans:2018qwy}
	\begin{eqnarray}\label{gm331}
		\left(\frac{g_{a\gamma \gamma}}{m_{a}}\right)\gtrsim 0.73\times \left(\frac{\alpha}{m_{e}^{2}}\right)=\frac{2.05\times 10^{-5} }{(\mathrm{eV})\cdot (\mathrm{GeV})}~.
	\end{eqnarray}
	Moreover, close to the resonance ($\omega_\textbf{p}\simeq \frac{\ell}{2}(m_a^2+\frac{1}{\ell^2})$ in which $\ell$ is defined below in Eq. \eqref{bg}) the ALP vacuum effect surpasses the virtual electron-positron effect in an even wider parameter region (see below).

	\textit{\it Spatially harmonic magnetic field profile.---}%
	We choose a spatial harmonic form for the magnetic field along the cavity axis, $x$-direction, with wavelength $2\pi\ell$
	\begin{eqnarray}\label{bg}
		\mathbf{B}(x)= \mathbf{B}_{0}\cos{\left(x/\ell\right)}~.
	\end{eqnarray}
	Inserting this magnetic field profile in the interaction Hamiltonian \eqref{hamiltonian12} and integrating over $\textbf{x}$, $\textbf{x}'$, and $\textbf{k} $ we find 
	\begin{equation}
		H_{\gamma B} = \sum_{s,s'\!,{\bf p}}\mathcal{F}^\mathbf{p}(\bm{\varepsilon}^{s'}\cdot\hat{\mathbf{b}})(\bm{\varepsilon}^{s}\cdot\hat{\mathbf{b}}) \,a_{s'}^{\dag}({\bf p}) a_s({\bf p})~.
		\label{hamil}
	\end{equation}
	Here, $\hat{\bf b}$ is a unit vector pointing in the direction of magnetic field and the interaction rate is complex $\mathcal{F}^\mathbf{p} \equiv \mathcal{F}_{r}^\mathbf{p} +i\mathcal{F}_{i}^\mathbf{p}$ with real and imaginary parts given by
	\begin{subequations}
		\begin{align}
			\mathcal{F}_{r}^\mathbf{p}&=G_{a}\int dk_{x}\frac{\omega_\mathbf{p}(p_{x}^2-k_{x}^2-m_{a}^2)\mathcal{P}(k_{x},p_x)}{(p_{x}^2-k_{x}^2-m_{a}^2)^2+(\omega_{\mathbf{p}}\Gamma_{\!B})^2}~,\label{Fr}\\
				\mathcal{F}_{i}^\mathbf{p}&=G_{a}\int dk_{x}\frac{\omega_{\mathbf{p}}^2\,\Gamma_{\!B}\,\mathcal{P}(k_{x},p_x)}{(p_{x}^2-k_{x}^2-m_{a}^2)^2+(\omega_{\mathbf{p}}\Gamma_{\!B})^2}~,\label{Fi}
			\end{align}
			\label{coupling}
	\end{subequations}
	where we have introduced $G_a\equiv g_{a\gamma\gamma}^2 B_0^2/2 \pi L_x$ and the profile function
		\begin{align}\label{Profile}
			\mathcal{P}(k_x,p_x)=&\frac{1}{4} \left[(-1)^{l_1}\left(\frac{\sin(\frac{L_x}{2}\Delta_1)}{\Delta_1}+\frac{\sin(\frac{L_x}{2}\Delta_2)}{\Delta_2}\right)\right.\nonumber\\
			-&\left.\left(\frac{\sin(\frac{L_x}{2}\Delta_3)}{\Delta_3}+\frac{\sin(\frac{L_x}{2}\Delta_4)}{\Delta_4}\right)\right]^2~,
		\end{align}
		in which $\Delta_1=k_x+p_x+\frac{1}{\ell}$, $\Delta_2=k_x+p_x-\frac{1}{\ell}$, $\Delta_3=k_x-p_x+\frac{1}{\ell}$, $\Delta_4=k_x-p_x-\frac{1}{\ell}$, and $l_1=p_xL_x/\pi$.
	The axion photon conversion  rate in the presence of a harmonic magnetic field (\ref{bg}) takes the following form (see Supplementary Material for details)
 \begin{eqnarray}\label{gammaB}
			\Gamma_B=\frac{g^2_{\gamma a} |\mathbf{B}_{0}|^2}{2\,L_x}  \mathcal{P}\left(k_x,\sqrt{k_{x}^{2}+m^{2}}\right) ~.
	\end{eqnarray}
On the other hand, in the presence of a uniform magnetic field the interaction rate of photons  is obtained as
	\beq
	\mathcal{F}^\mathbf{p} = 
	- \frac{ g_{a\gamma\gamma}^{2}B_{0}^{2} \omega_{\mathbf{p}}}{2} 
	\left(\frac{m_{a}^{2} +i\omega_{\mathbf{p}}\Gamma_{\!B_{0}}}{m_{a}^4+(\omega_{\mathbf{p}}\Gamma_{\!B_{0}})^{2}}\right),
	\eeq
	where  the conversion rate is  $\Gamma_{\!B_{0}}=g_{a\gamma\gamma}^{2}  B_{0}^2 L_x/8$.
	
	In the rest of this Letter, we employ the Hamiltonian \eqref{hamil} to study the dynamics of photon polarization of the cavity modes and determine the minimum detectable coupling rates $\mathcal{F}$ for feasible experimental parameters.
	
	\textit{\it Cavity detection scheme.---}%
	In order to probe the virtual ALPs we propose a detection scheme based on high-precision phase measurement on the field of a high-finesse cavity.
	A Fabry-Perot optical cavity of length $L_x$ is placed inside a nonuniform magnetic field perpendicularly applied to the cavity axis [see Fig.~\ref{fig:scheme}].
	By assuming a one-dimensional (1D) model, the cavity longitudinal mode frequencies are found as $\omega_n=n\pi c/L_x$ with $n=1,2,3,\cdots$. The cavity modes are separated by the free spectral range (FSR) $\Delta_\mathrm{FSR}=\pi c/L_x$ and at each frequency the cavity supports two orthogonal polarization modes $\{\bm\varepsilon^1,\bm\varepsilon^2\}$.
	At each run of the experiment only one of the cavity modes is selected by coherently pumping at a laser frequency $\omega_L$---on resonance with the mode.
	Scanning over a given mass range can be done by changing frequency of the laser.
	The accumulated optical field inside the cavity thus becomes a coherent state $|\alpha\rangle$ with a well-defined amplitude and phase $\alpha = |\alpha|e^{i\phi}$.
	According to our theory, in the absence of a magnetic field no axion-photon interaction is expected. One, thus, expects a trivial time evolution for the field with the dynamical phase determined by the pump frequency $|\alpha\rangle(t)=|\alpha e^{-i\omega_Lt}\rangle$.
	Nevertheless, in the presence of a magnetic field the axion-photon interaction triggers the polarization-flip processes. One of the polarizations, $\bm\varepsilon^1$, is aligned with the magnetic field direction ($z$-axis) to have interaction with ALPs according to Eq. \eqref{hamil}. The other one, $\bm\varepsilon^2$, ought to be perpendicular to the magnetic field direction (e.g. $y$-axis) in order to remain intact and be chosen as a reference polarization. The benefit of this configuration is that both polarizations sense the same noise inside the cavity.
	
	The multi-mode Hamiltonian in Eq.~(\ref{hamil}) simplifies to $H = \sum_s\!\big(\omega_L a_s^\dag a_s +\sum_{s'} \mathcal{F} a_s^\dag a_{s'}\big)$ that describes coherent evolution of the two polarization modes with the selected frequency. Note that we have dropped the $\mathbf{p}$ dependency in the Hamiltonian for the convenience.
	$\mathcal{F}=\mathcal{F}_r+i\mathcal{F}_i$ is a complex number given in Eqs.~\eqref{coupling}, therefore, the Hamiltonian is not Hermitian. Its imaginary part modifies the decay rate of the cavity mode. We employ the quantum Langevin equation formalism to study the dynamics of the cavity modes~\cite{Abdi2012, Abdi2019}. The loss and noise processes are introduced by the input-output theory~\cite{Walls2007}
	\begin{subequations}
		\begin{align}
			\dot{a}_1 &= -(\tilde\kappa +i\tilde\omega)a_1-i\mathcal{F}a_2 +\sqrt{2\kappa}\ a_1^{\rm in},\label{Langevin-a1} \\
			\dot{a}_2 &= -(\kappa +i\omega)a_2-i\mathcal{F}a_1+\sqrt{2\kappa}\ a_2^{\rm in},\label{Langevin-a2}
		\end{align}
	\end{subequations}
	where $a_j^{\rm in}$ are the input noise operators. Here, $\tilde\kappa \equiv \kappa -\mathcal{F}_i$ is the modified cavity decay rate, while $\tilde\omega \equiv \omega_L +\mathcal{F}_r$ is the axion-dispersed resonance frequency of the driven cavity mode. 
	
	The occurrence of the photon polarization-flip is then detected by measuring and comparing the phase of the cavity output fields using $a_2$ as a reference [see Fig.~\ref{fig:scheme}]. In the absence of interaction the phase deviation $\delta\phi$ is only caused by QED process which is the background noise. However, when the axion-induced dispersion gets activated by the static magnetic field, we anticipate a deviation in the phase of the normal modes as: $\delta\phi_{1} =\mathcal{F}_r\tau$ and $\delta\phi_{2} =0$ for optical pulses of duration  $\tau$.
	We have neglected differences in the decay rate of the modes as it becomes negligible even for the highest available cavity quality factors. 
	The accumulated phase $\delta\phi_1$ can be enhanced with a longer measurement time. However, the maximum available $\tau$ is limited by the cavity lifetime $\tau_{\rm max}\approx 1/\kappa$ which is attainable for sufficiently long laser pulses.
	The exclusion or detection of ALPs at a given mass thus boils down to the precision phase mismatch measurement in the optical cavity output.
	Thanks to the recent advances in the phase measurement techniques, the phase sensitivity is only limited by shot-noise in a broad frequency range; from terahertz to optical~\cite{Echternach2013, Manceau2017}. Therefore, the smallest detectable phase deviation for a coherent state is
	$\delta\phi_1 = N_{\rm ph}^{-1/2}$,
	where $N_{\rm ph}$ is the average number of photons in the optical pulse. Therefore, the minimum detectable coupling rate for a single run of experiment is $\delta\mathcal{F}_r = \kappa/N_{\rm ph}^{1/2}$.
	In practice the experiment runs several times at time intervals determined by the optical pulse duration.
	By repeating the procedure for $N_{\rm exp}$ times the precision gets enhanced by reducing the mean-standard-deviation.
	The phase sensitivity is increased further by employing squeezed light pumps. Currently, quadrature squeezing as high as 15~dB is available and higher values are within  reach~\cite{Vahlbruch2016}. The reduced phase quadrature extension is parametrized by the factor $e^{-r}$ through the squeezing parameter $r$, e.g. $r=0$ for a coherent light, while $r\approx 1.73$ indicates a $15$~dB squeezing.
	In terms of the cavity length $L_x$ and finesse $F$ we find the following expression for the $\mathcal{F}_r$ resolution
	\begin{equation}
		\delta\mathcal{F}_r = \frac{e^{-r}}{\sqrt{N_{\rm ph}N_{\rm exp}}}\cdot\frac{\pi c}{FL_x}~.
		\label{resolution}
	\end{equation}
	\begin{figure}[t]
		\centering
		\textbf{}\includegraphics[width=\columnwidth]{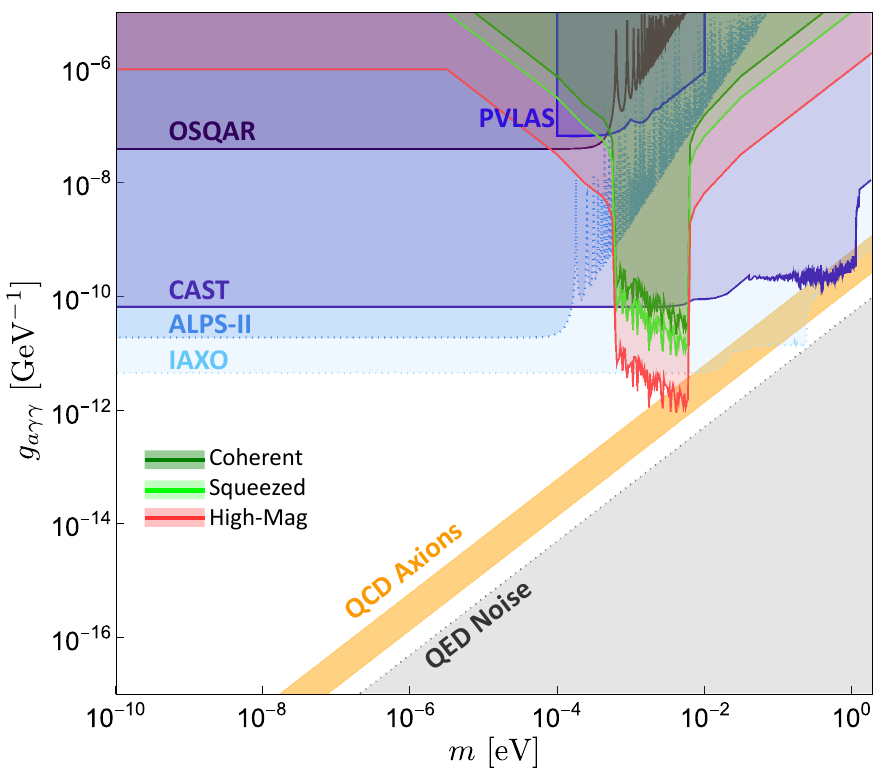}
		\caption{Exclusion areas in the plane of ALP mass and axion-photon coupling constant  for three different scenarios discussed in the text. The gray shade is the region where QED-induced polarization-flip dominates that of axion-induced processes. The dotted line corresponds to $\mathcal{F}_r=3\mathcal{G}^\textbf{p}$.}
		\label{fig:exclusion}
	\end{figure}
	\textit{\it Exclusion areas.---}%
	We now discuss the feasibility of the experimental implementation of our scheme and the areas of $g_{a\gamma\gamma}$-$m_a$ that can be essentially excluded. From Eq.~\eqref{coupling} one expects a maximized coupling rate at the resonance $\omega_L=\frac{\ell}{2}(m_a^2+\frac{1}{\ell^2})$ and thus the exclusion extends to the smaller $g_{a\gamma\gamma}$ values, which is lower-bounded by the finite cavity linewidth $\kappa$. In particular, the high precision phase measurement techniques at optical frequencies allow for excluding by our setup.
	In order to implement the setup as a tabletop experiment we consider an optical cavity of $L_x\approx 1~m$~\cite{BiRefCav:2019,Ring:2020,fujimoto2021dark} and a magnetic field of sine profile with the characteristic length of $\ell=\frac{L_x}{10\pi}$.
	We analytically evaluate $\mathcal{F}_r$ using Eq.~\eqref{analytic} for a wide range of $m_a$ and $g_{a\gamma\gamma}$ and compare it to $\delta\mathcal{F}_r$ obtained from \eqref{resolution}.
	For a cavity finesse $F=10^5$, average pulse photon number $N_{\rm ph}=10^{14}$, and a moderate magnetic field of $B_0=10$~T~\cite{Anastassopoulos:2017ftl} one already calculates $\tau\simeq10^{-4}$ s and $\delta\mathcal{F}_r \simeq 10^{-3}$~Hz. This gets enhanced by three orders of magnitude with a million time repetition of the experiment $N_{\rm exp}=10^6$ and improves by another order of magnitude when a $15$~dB squeezed light is employed, dragging the resolution down to $10^{-7}$~Hz. The state-of-the-art sources of high magnetic field are nowadays able to reach at least ten times larger values~\cite{Battesti2018}, which accounts for an order of magnitude improvement in the exclusion region. Considering the optical pulse duration and the required time for homodyne detection, each run of the experiment takes about $\tau_{exp}\sim\tau$. In our scenario using the above-mentioned number of repetitions $N_{exp}=10^6$ and a cavity with the length and finesse of $F=10^5$ and $L_x=1$ m, respectively, examining each ALP mass needs about $N_{exp}\tau_{exp}\approx17$ minutes. Therefore, by our scheme one examines about 2500 ALP masses in a period of one-month experiment.
	
	Fig.~\ref{fig:exclusion} shows the areas that can be excluded by our scheme at three different conditions:
	(i) coherent light pump and moderate magnetic fields ($B_0=10$~T).
	(ii) squeezed state ($r=1.73$) and moderate magnetic field.
	(iii) squeezed state and ultrahigh magnetic field ($B_0=100$~T).
	In the figure only resonance points $m\approx n\Delta_{\rm FSR}~(n=1,2,3,\cdots)$ are illustrated considering the finite cavity linewidth $\kappa$. Between these cavity resonances the detection falls as a Lorentzian function. Thanks to the small FSR value ($\Delta_{\rm FSR}\approx 6.2\times 10^{-7}$~eV) the resonance points get tightly positioned, covering a large amount of possible ALPs masses. Moreover, practical considerations of optical cavities enforce us to use laser frequencies of the range $0.03\;\text{eV}\leq\omega_L\leq3\;\text{eV}$ and due to the resonance condition $\omega_\textbf{p}\simeq \frac{\ell}{2}(m_a^2+\frac{1}{\ell^2})$ the best performance of the scheme is limited to a mass range of $6\times10^{-4}\;\text{eV}\leq m_a\leq6\times10^{-3}\;\text{eV}$ as shown in the figure. As mentioned above, about 2500 ALP masses in this region could be examined by a one-month experiment. Increasing and decreasing $\ell$ shifts the exclusion region to right and left, respectively, and the coverage of the experiment further extends.  Remarkably, the sensitivity in our proposed experiment goes  beyond  the projected sensitivity of IAXO \cite{Armengaud:2019uso} and reaches to the QCD axion parameter region.
	
	In the case of a uniform magnetic field the phase shift is easily attained analytically
	\begin{equation} \label{PhaseConstantField}
		\delta\phi_1=\frac{g_{a\gamma\gamma}^{2}B_{0}^{2} m_{a}^{2}}{m_{a}^{4}+(\omega_\textbf{p}\Gamma_{\!B_{0}})^{2}}~\frac{\omega_\textbf{p}\tau}{2}~.
	\end{equation}
	For the parameters considered in this work one finds that the  constant magnetic field gives a much weaker constraint. For instance, considering $g_{a\gamma\gamma}=10^{-10}\;\text{GeV}^{-1}$, $B_0=10\;\text{T}$, and $m_a=10^{-4}\;\text{eV}$ near the resonance point $\omega_{\mathbf{p}}\approx0.08\;\text{eV}$ the phase deviation of $\bm\varepsilon^1$ polarization photons due to the constant magnetic field based on Eq. \eqref{PhaseConstantField} is $\delta\phi_1\approx\;2.5\times10^{-16}$. However, exploiting the benefit of a harmonic magnetic field based on Eq. \eqref{analytic} in the same conditions results in a phase deviation of $\delta\phi_1\approx\;4.6\times10^{-11}$ which shows about 5 orders of magnitude improvement in the phase deviation inside the cavity.

	\textit{\it Discussion.---}%
	As a competing effect, the QED-induced photon scattering in a background magnetic field is one of the most prohibitive effects whose contribution should be carefully taken into account. We now investigate this effect and show that indeed there are interesting parameter regions where the ALP becomes the prominent effect.
	
	Taking the effective Euler-Heisenberg Lagrangian at one-loop calculation~\cite{Euler1936, Heisenberg1936, Dicus1998,Dunne:2012hp} and following the straightforward calculations one finds $H_{\rm QED} = \sum\mathcal{G}^\mathbf{p}(\bm{\varepsilon}^{s'}\!\cdot\hat{\mathbf{b}})(\bm{\varepsilon}^{s}\!\cdot\hat{\mathbf{b}}) \,a_{s'}^{\dag}({\bf p}) a_s({\bf p})$  for the photon interaction~\cite{Shakeri:2017knk,Shakeri:2017iph}. This has the same form as Eq.~\eqref{hamil} though with a different coupling rate that for a harmonic magnetic field under the forward scattering condition is
	\beq
	\mathcal{G}^{\bf p}\equiv-\frac{\alpha^{2}B_{0}^{2}\omega_{\mathbf{p}}}{30m^{4}_{e}}\Big(1+\text{sinc}\big(\frac{L_x}{\ell} \big)\Big)~.
	\eeq
	The case of uniform magnetic field is retrieved at the limit of $\ell\gg L_x$.
	
	The phase shift contribution from QED scattering is similarly obtained as $\delta\phi_{1,\textsc{QED}}=\mathcal{G}^\textbf{p}\tau$. The region at which ALP effect is dominated by the QED scattering is separated by a gray shade in Fig.~\ref{fig:exclusion}.
	Remarkably, in most  exclusion regions our scheme is immune from QED effects. 
	
	In summary, we have proposed a new detection method for ALPs based on the high-precision phase measurement.  The phase difference between two polarization modes of photons can be induced by off-shell ALPs  during the forward scattering from a non-uniform magnetic field.  Our scheme can potentially impact significantly on the discovery of ALPs as it covers a mostly overlooked and unexplored parameter region. Our setup paves the way for searching in a mass range reaching the QCD axion parameter region.

	\begin{acknowledgements}
		\textit{\it Acknowledgements.---}%
		MZ acknowledges financial support by the University of Padova under the “MSCA Seal of Excellence @UniPD programme.
		MA acknowledges support by STDPO and IUT through SBNHPCC.
		DJEM was supported
		by the Alexander von Humboldt Foundation and the German Federal Ministry of Education and Research, and is now supported by an Ernest Rutherford Fellowship of the Science and Technologies Facilities Council (UK).
		SM acknowledges partial financial support by ASI Grant No. 2016-24-H.0.
		SS thanks to Horst Fisher and Marc Schumann for the support during ``15th Patras Workshop on Axions, WIMPs and WISPs" at University of Freiburg. SS is also grateful to Georg-August Universit\"at of G\"ottingen for  kind hospitality where this work was in progress.
	\end{acknowledgements}
	
	\bibliography{references.bib}

	\clearpage
	\newpage
	\maketitle
	\onecolumngrid
	\begin{center}
		\textbf{\large Probing Virtual Axion-Like Particles by Precision Phase Measurements} \\ 
		\vspace{0.05in}
		{ \it \large Supplementary Material}\\ 
		\vspace{0.05in}
		{}

	\end{center}
	\setcounter{equation}{0}
	\setcounter{figure}{0}
	\setcounter{table}{0}
	\setcounter{section}{1}
	\renewcommand{\theequation}{S\arabic{equation}}
	\renewcommand{\thefigure}{S\arabic{figure}}
	\renewcommand{\thetable}{S\arabic{table}}
	\newcommand\ptwiddle[1]{\mathord{\mathop{#1}\limits^{\scriptscriptstyle(\sim)}}}

\begin{figure} [t]
		\centering  \label{plotscattering}
		\includegraphics[width=3.5in]{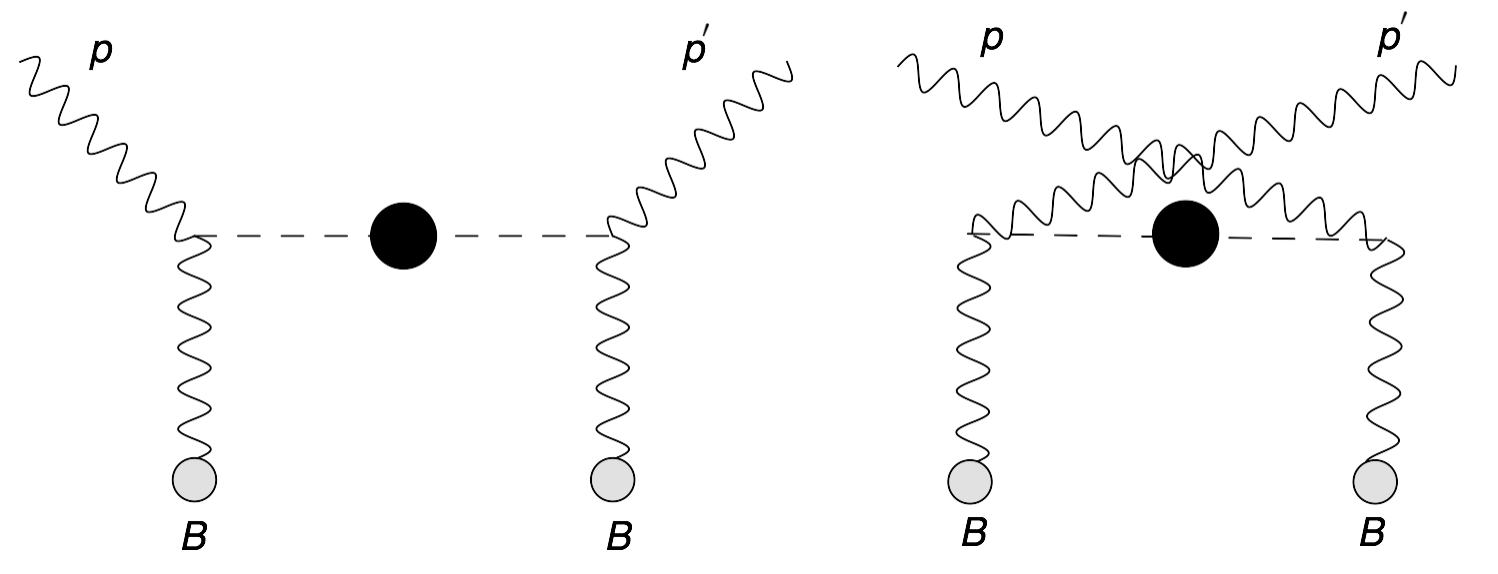}
		\caption{Photon scattering from an external  magnetic field. Dashed lines with a solid ball at the center  represents the effective axion  propagator in the presence of a non-uniform  magnetic field.}\label{a1}
	\end{figure}

\begin{figure} [t]
		\centering 
		\includegraphics[width=6in]{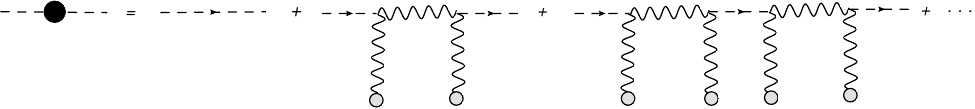}
		\caption{The effective axion propagator in the presence of an external magnetic field which is obtained from   a sum of the full series of diagrams with external field insertion.}\label{a2}
	\end{figure}

	In this Supplementary Material, we provide further details about the conventional quantum field theory calculations
	of the second-order interaction Hamiltonian   (\ref{hamiltonian12}) and the axion photon conversion  rate $\Gamma_{B}$ in the presence of a harmonic magnetic field (\ref{gammaB}).
	The second-order S-matrix element corresponding to
	the Feynman diagrams in Fig. \ref{a1}
	is
	\begin{align} 
		S^{(2)}(\gamma_s \rightarrow\gamma_{s'})
		=-\frac{i}{8}g^2_{a\gamma\gamma}\int\! d^4x' d^4x\hspace{1mm}\epsilon^{\mu\nu\alpha\beta}\epsilon^{\mu'\!\nu'\!\alpha'\!\beta'} \bar{F}_{\alpha\beta} (x')\bar{F}_{\alpha'\beta'} (x)   D(x'-x)\left [\partial_{\mu}A^{-}_{\nu}(x') \partial_{\mu'}A^{+}_{\nu'}(x) + \partial_{\mu'}A^{-}_{\nu'}(x) \partial_{\mu}A^{+}_{\nu}(x')\right]~,
	\end{align}
	using this, the effective second-order Hamiltonian $H_{\gamma B}$ is defined by
	\begin{align} \label{HgammaB}
		H_{\gamma  B}=\frac{1}{8}g^2_{a\gamma\gamma}\int\ d^4x' d^3x\,\epsilon^{\mu\nu\alpha\beta}\epsilon^{\mu'\nu'\alpha'\!\beta'}\bar{F}_{\alpha\beta} (x')\bar{F}_{\alpha'\beta'} (x') 
		D(x'-x)\left [\partial_{\mu}A^{-}_{\nu}(x') \partial_{\mu'}A^{+}_{\nu'}(x) 
		+ \partial_{\mu'}A^{-}_{\nu'}(x) \partial_{\mu}A^{+}_{\nu}(x')\right]~,
	\end{align}
	where $D(x'-x)$ is the effective ALPs propagator,  $ A^{+}_{\mu} \,(A^{-}_{\mu})$ is  linear in  annihilation (creation) operator $a_s$ ($a^{\dag}_s$) of photons. By quantizating the photon field inside the cavity~\cite{kakazu1994quantization} and considering boundary conditions on $x=\pm \frac{L_x}{2}$, the Fourier transform of the photon field $A_{\nu}(x)$ is
\begin{eqnarray}    \label{quantumA}
			A_\nu(x)=A_\nu^+(x) + A_\nu^-(x)=\sum_{s}\sum_{\textbf{p}}\frac{1}{\sqrt{\omega_{\textbf{p}} V}}\left[ i\sin(p_{x} x+\frac{ p_xL_x}{2})\,a_s(\textbf{p})\,e^{i p_y y}e^{i p_z z}e^{-i\omega_{\textbf{p}}t}\,\varepsilon^s_\nu(\textbf{p})+ \text{h.c.}  \right]~,
		\end{eqnarray}
		where $\textbf{p}=\left(p_{x},p_{y},p_{z}\right)=\left(\frac{\pi l_1}{L_x},\frac{2\pi l_2}{L_y},\frac{2\pi l_3}{L_z}\right)$ with $l_2,l_3=0,\pm1,\pm2,  \cdot\cdot\cdot$ and $l_1=0, 1, 2, \cdot\cdot\cdot$, $V$ is the cavity volume, and  the summation convention implied over  photon polarization states $s=1,2$ and different energy modes of the cavity photons. Plugging \eqref{quantumA} into \eqref{HgammaB} we get
		\bea 
		H_{\gamma B}(t)&=&\frac{g^2_{a\gamma\gamma}}{2 V}
		\,\sum_{s,s'}\sum_{\textbf{p} ,\textbf{p}'}\sqrt{\omega_\mathbf{p}\omega_\mathbf{\textbf{p}'}}
		\int dt'dx'd^2 \textbf{x}'_\bot \,dx\, d^2 \textbf{x}_\bot\,\frac{d^4k}{(2\pi)^4}\,D(k)
		\big[\bm{\varepsilon}^{s'*}(\textbf{p}')\cdot\textbf{B}(x')\big]\big[\bm{\varepsilon}^{s}(\textbf{p})\cdot\textbf{B}(x)\big]
		\nonumber \\ &\times &
		a_{s'}^{\dagger}(\textbf{p}')a_{s}(\textbf{p}) \left[\sin(p'_{x} x+\frac{p'_x L_x}{2})\sin(p_{x} x'+\frac{p_x L_x}{2})
		e^{-i\textbf{k}\cdot \textbf{x}' }
		e^{-i \textbf{p}'_{\bot}\cdot   \textbf{x}_\bot}
		e^{i\textbf{k}\cdot \textbf{x} }
		e^{i \textbf{p}_\bot\cdot   \textbf{x}'_\bot}
		e^{-i(-k^0+\omega_{\mathbf{p}})t'}
		e^{-i(k^0-\omega_{\mathbf{p}'})t}
		\right.\nonumber\\
		&+& \left. \sin(p'_{x} x'+\frac{p'_x L_x}{2})\sin(p_{x} x+\frac{p_x L_x}{2})
		e^{-i\textbf{k}\cdot \textbf{x}' }
		e^{-i \textbf{p}'_{\bot}\cdot   \textbf{x}'_\bot}
		e^{i\textbf{k}\cdot \textbf{x} }
		e^{i \textbf{p}_\bot\cdot   \textbf{x}_\bot}
		e^{-i(-k^0-\omega_{\textbf{p}'})t'}
		e^{-i(k^0+\omega_{\textbf{p}})t}\right]~,
		\eea
	in which $D(k)$ is the effective ALP propagator in the  momentum space which is given by
	\bea
	iD(k)&=&\frac{i}{k^2-m_0^2}+\frac{i}{k^2-m_0^2}(-i\Pi_B(k))\frac{i}{k^2-m_0^2}+\cdot\cdot\cdot
	= \frac{i}{k^2-m_{a}^2-i\text{Im} \Pi_B(k)}~,
	\eea
	where $-i\Pi_B(k)$ denotes the sum of all one-particle irreducible ($1\text{PI}$) diagrams including  diagrams with two external magnetic field lines [Fig. \ref{a2}].  The imaginary part of $\Pi_B$ can be identified as the conversion rate of axion into photon in the presence of the background magnetic field
 \begin{equation}
			\mathrm{Im}\Pi_B=-k^0\,\Gamma_{B}~.
	\end{equation}
	Imposing the forward scattering condition ${\bf p}={\bf p'}$, the interaction Hamiltonian is represented as 
\begin{eqnarray} \label{hamiltonian1}
			H_{\gamma B}&=&\frac{g^2_{a\gamma\gamma}}{2V}\int dx'd^2 \textbf{x}'_\bot \,dx\, d^2 \textbf{x}_\bot\,\frac{d^4k}{(2\pi)^4}\,
			\sum_{s,s',\mathbf{p}}
			\frac{\omega_{\mathbf{p}} a_{s'}^{\dag}({\bf p}) a_s({\bf p})}{k^2-m_{a}^2+ik^0\Gamma_B}
			\big[\bm{\varepsilon}^{s'*}(\textbf{p}')\cdot\textbf{B}(x')\big]\big[\bm{\varepsilon}^{s}(\textbf{p})\cdot\textbf{B}(x)\big]\nonumber\\
			&\times&\sin(p_{x} x'+\frac{p_x L_x}{2})\sin(p_{x} x+\frac{p_x L_x}{2})\left(e^{-i\textbf{k}\cdot \textbf{x}' }
			e^{-i \textbf{p}_{\bot}\cdot   \textbf{x}_\bot}
			e^{i\textbf{k}\cdot \textbf{x} }
			e^{i \textbf{p}_\bot\cdot   \textbf{x}'_\bot}+
			e^{-i\textbf{k}\cdot \textbf{x}' }
			e^{-i \textbf{p}_{\bot}\cdot   \textbf{x}'_\bot}
			e^{i\textbf{k}\cdot \textbf{x} }
			e^{i \textbf{p}_\bot\cdot   \textbf{x}_\bot}\right)~.
	\end{eqnarray}
In the following, we calculate  $\Gamma_B$ for a given profile of the magnetic field. The amplitude for the axion-photon conversion process in the presence of the magnetic field is given by
	\bea
	\mathcal{T}_{fi}&=&\left<a,\mathbf{ k}\right|\frac{g_{\gamma a}}{2}\int d^4x\, \epsilon^{\mu\nu\alpha\beta} \phi(x)\bar{F}_{\alpha\beta}(x)\partial_{\mu}A_{\nu}(x)\left|\gamma,\mathbf{p}\right>\\ \nonumber 
	&=&-\frac{ig_{\gamma a} \sqrt{\omega_{\mathbf{p}}}}{ V\sqrt{2k^0}}
		\int_{-T/2}^{T/2} dt e^{-i(\omega_{\mathbf{p}}-\omega_{\mathbf{k}}) t}
		\int_{V} d^3\textbf{x}\,   e^{i(\mathbf{p}_\bot-\mathbf{k}_\bot)\cdot \mathbf{x}_\bot}\,e^{-ik_x x}\sin(p_x x+\frac{p_x L_x}{2})\,\bm{\varepsilon}^{s}(\mathbf{ p})\cdot\mathbf{B}(x)~.
	\eea
	Now, for a spatially harmonic magnetic field $\mathbf{B}(x)= \mathbf{B}_{0}\,\cos(\frac{x} {\ell})$, the conversion amplitude transforms to 
\bea
	\mathcal{T}_{fi}&=&-\frac{ig_{\gamma a} \sqrt{\omega_{\mathbf{p}}}\bm{\varepsilon}^{s}(\mathbf{ p})\cdot\mathbf{B}_{0}}{V \sqrt{2k^0}}\int_{-T/2}^{T/2} dt\, e^{-i(\omega_{\mathbf{p}}-\omega_{\mathbf{k}}) t}\int_{S} d^2\textbf{x}_{\bot}e^{i(\mathbf{p}_{\bot}-\mathbf{k}_{\bot})\cdot \mathbf{x}_{\bot}}\int_{-L_x/2}^{L_x/2} dx\,   e^{-ik_{x}x}\sin(p_xx+\frac{p_x L_x}{2})\cos(\frac{x} {\ell})~.
\eea
Here, ${\bot}$ shows the  components in $yz$-plane perpendicular to the direction of the magnetic field oscillations. Note that the magnetic field oscillation length  $\ell$   is different from the cavity length L and the cavity volume is $V=LS$. The conversion probability per unit time is given by
\bea
		w=\frac{|\mathcal{T}_{fi}|^{2}}{T}&=&\frac{g^2_{\gamma a}  \omega_{\mathbf{p}}}{ 2\,L_x V k^0} |\mathbf{B}_{0}|^2(2\pi)^3 \delta(\omega_{\mathbf{p}}-\omega_{\mathbf{k}})\delta^2(\mathbf{p}_{\bot}-\mathbf{k}_{\bot}) \mathcal{P}(k_x,p_x)~,
\eea
where $\mathcal{P}(k_x,p_x)$ is the profile function.
 \bea\label{Gmmm}
		\Gamma_B=\int w \frac{Vd^3p}{(2\pi)^3}=\frac{g^2_{\gamma a} |\mathbf{B}_{0}|^2}{2\,L_x} \int dp_{x} \frac{\omega_{\mathbf{p}}}{k^0}\delta(\omega_{\mathbf{p}}-k^0)  \mathcal{P}(k_x,p_x) ~,
		\eea
	applying well-known properties of the Dirac-delta function we find
 \bea\label{Gmmm}
		\Gamma_B=\frac{g^2_{\gamma a} |\mathbf{B}_{0}|^2}{2\,L_x}  \mathcal{P}\left(k_x,\sqrt{k_{x}^{2}+m^{2}}\right) ~,
\eea
	In the presence of a constant magnetic field $\mathbf{B}= \mathbf{B}_{0}$, the momentum transfer $q_{x}$ tends to zero and the decay rate is given by
\begin{eqnarray}
			\Gamma_{B_{0}}=\frac{g^2_{\gamma a} |\mathbf{B}_{0}|^2}{2\,L_x}\left(\frac{\sin\left(\frac{L_x}{2}(k_x-\sqrt{k_{x}^{2}+m^{2}})\right)}{k_x-\sqrt{k_{x}^{2}+m^{2}}}\right)^2
	\end{eqnarray}
	by assuming $\mathbf{p}_{\bot}=0$, \emph{we recover well-known result of the axion photon conversion rate} \cite{Sikivie1983,Ioannisian:2017srr,Tam:2011kw} as 
\begin{eqnarray}\label{ga44}
			\Gamma_{B_{0}}=\frac{g_{\gamma a}^{2}   |\mathbf{B}_{0}|^2 }{2L_x} \left( \frac{\sin{(q_{x}L_x/2)}}{q_{x}} \right)^{2},
	\end{eqnarray}
where $q_x=k_x-\sqrt{k_x^2+m^2}$ and in the case of $q_{x} L_x\ll1$, on can approximate Eq. (\ref{ga44}) as 
\begin{eqnarray}
			\Gamma_{B_{0}}=\frac{g_{\gamma a}^{2}   |\mathbf{B}_{0}|^2 L_x}{8} 
\end{eqnarray}

	\newpage
	
	\section{The Enhancement Effect of Profile Function}
	
\begin{figure*}[h]
		\centering
		\captionsetup[subfigure]{oneside,margin={.63cm,0cm}}
		\begin{subfigure}{.48\textwidth}
			\hspace*{-.3cm}
			\includegraphics[width=\textwidth]{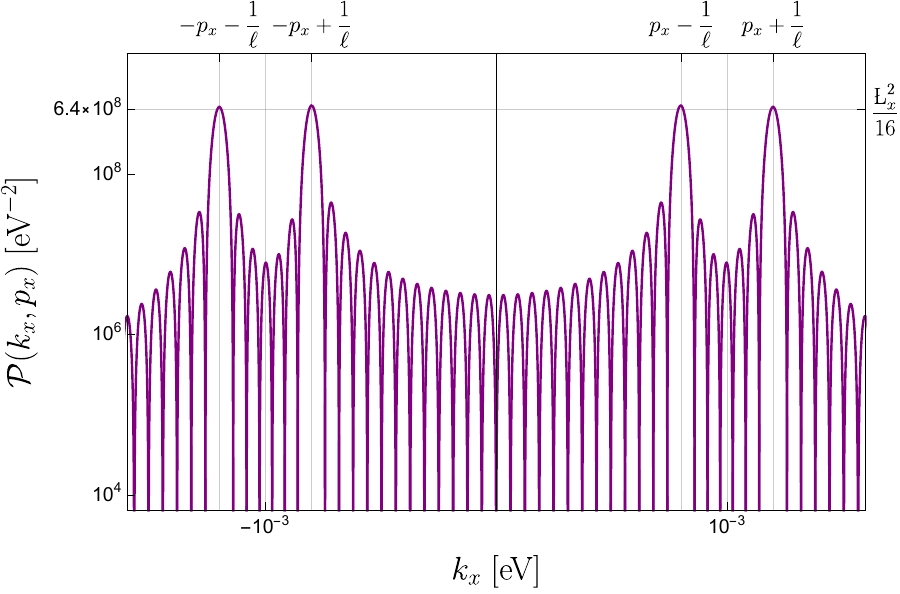}
			\caption{}
			\label{profile1:fig}
		\end{subfigure}
		\quad
		\captionsetup[subfigure]{oneside,margin={.9cm,0cm}}
		\begin{subfigure}{.48\textwidth}
			\includegraphics[width=\textwidth]{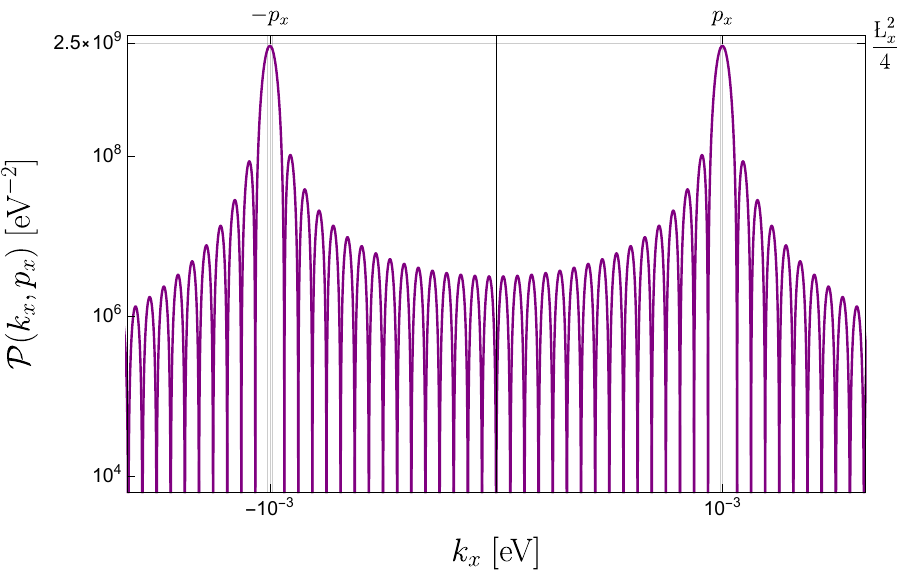}
			\caption{}
			\label{profile2:fig}
		\end{subfigure}
		\caption{ \small (a) The profile function of Eq. \eqref{Profile} for a 2 cm cavity which is pumped by $p_x=10^{-3}$ eV photons and exposed to a magnetic field with $\ell=0.05\;L_x$. There are four distinct peaks on $\pm p_x\pm\frac{1}{\ell}$.  (b) The same profile function with $\ell=L_x$.}  \label{profile}
	\end{figure*}

	As it can be seen from Eq. \eqref{Profile}, the profile function has four peaks which cause an enhancement in the detection. Fig. \ref{profile1:fig} is a depiction of  this profile function in which we have chosen $L_x=2$ cm, $\ell=0.02\;L_x$, and $p_x=10^{-3}$ eV. By increasing $\ell$, each pair of peaks at the right- and left-hand sides of Fig. \ref{profile1:fig} further approach to $p_x$ and $-p_x$, respectively. Eventually, as depicted in Fig. \ref{profile2:fig} for larger values of $\ell$ each pair of peaks merge together and create two peaks at $p_x$ and $-p_x$.
	
\begin{figure*}[h]
		\centering
		\captionsetup[subfigure]{oneside,margin={1.2cm,0cm}}
		\begin{subfigure}{.48\textwidth}
			\hspace*{-.3cm}
			\includegraphics[width=\textwidth]{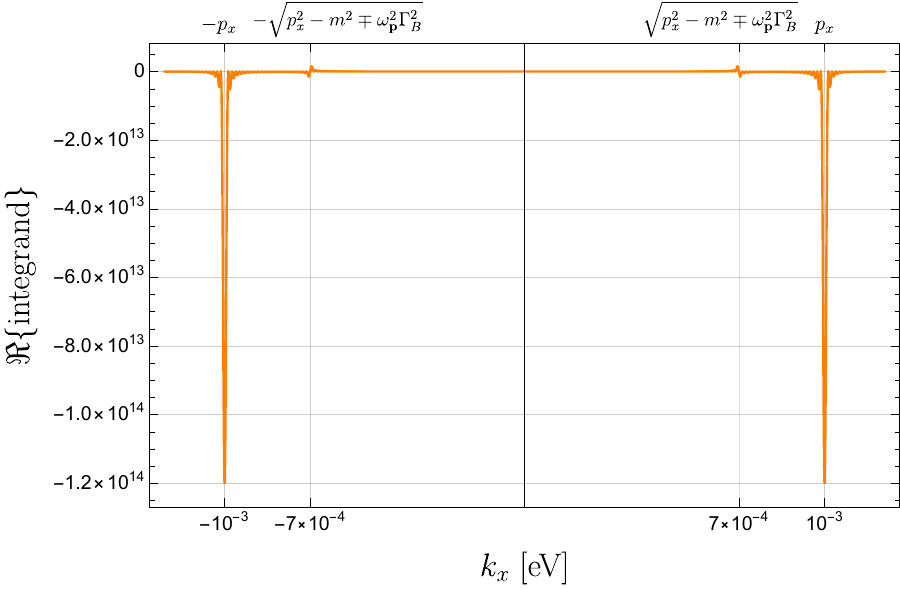}
			\caption{}
			\label{profile-re:fig}
		\end{subfigure}
		\quad
		\captionsetup[subfigure]{oneside,margin={1.41cm,0cm}}
		\begin{subfigure}{.48\textwidth}
			\includegraphics[width=\textwidth]{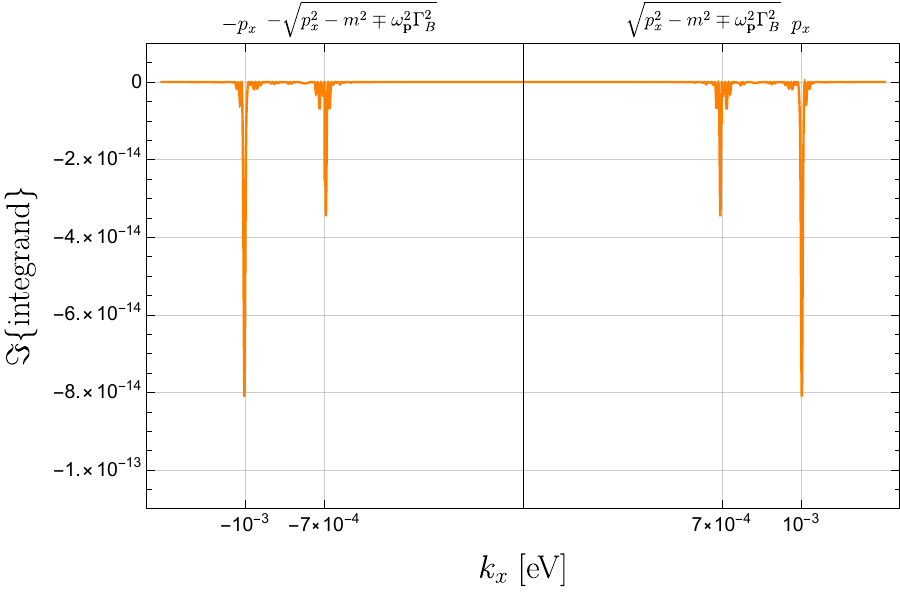}
			\caption{}
			\label{profile-im:fig}
		\end{subfigure}
		\caption{ \small (a) The integrand of Eq. \eqref{Fr} for a 10 cm cavity which is pumped by $p_x=10^{-3}$ eV photons and exposed to a magnetic field with $\ell=L_x$. The axion mass and the coupling constant are assumed to be $m=7\times10^{-4}$ eV and $g=10^{-10}\;\text{GeV}^{-1}$, respectively. (b) The integrand of Eq. \eqref{Fi} for the same scenario.}  \label{integrand}
	\end{figure*}

	The integrand function in Eq.~\eqref{coupling} has a propagator part and a profile function part. The propagator has four peaks at $\pm\sqrt{p_x^2-m^2\mp\omega^2_\textbf{p}\Gamma_B^2}$ and the profile also has four peaks at $\pm p_x\pm\frac{1}{\ell}$. The advantage of the profile peaks compared to the propagator ones can be seen in Fig. \ref{integrand} in which the integrand is illustrated for $m=7\times10^{-4}$ eV, $p_x=10^{-3}$ eV, $g=10^{-10}\;\text{GeV}^{-1}$, $L_x=10$ cm, and $\ell=L_x$. 
	
\begin{figure*}[h]
		\centering
		\captionsetup[subfigure]{oneside,margin={.8cm,0cm}}
		\begin{subfigure}{.48\textwidth}
			\hspace*{-.3cm}
			\includegraphics[width=\textwidth]{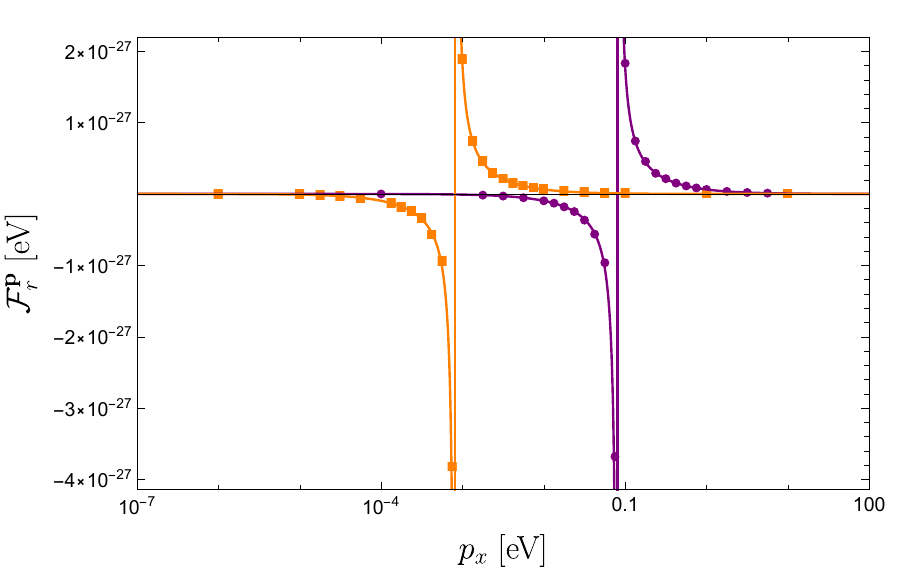}
			\caption{}
			\label{profile-re:fig}
		\end{subfigure}
		\quad
		\captionsetup[subfigure]{oneside,margin={1.39cm,0cm}}
		\begin{subfigure}{.48\textwidth}
			\includegraphics[width=\textwidth]{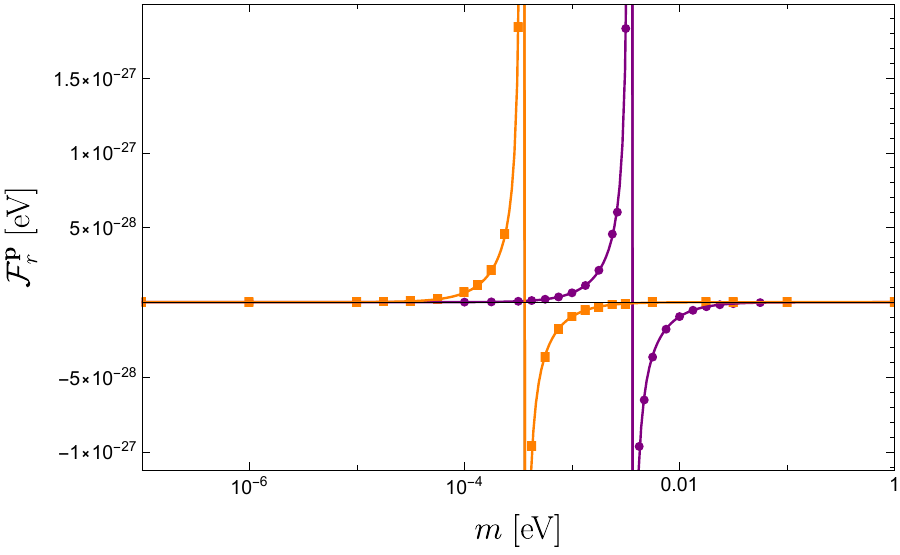}
			\caption{}
			\label{profile-im:fig}
		\end{subfigure}
		\caption{ \small A comparison between numeric and analytic evaluation of Eq.~\eqref{coupling}. Solid lines correspond to the analytic evaluation using \eqref{analytic} and the dots illustrate the numerical value. The cavity length is 1 m, the magnetic field has $\ell=L_x/10\pi$, and the coupling constant assumed to be $g_{a\gamma\gamma}=10^{-10}\;\text{GeV}^{-1}$. (a) The purple plot corresponds to the ALP mass of $m=10^{-3}$ eV and the orange one is for $m=10^{-4}$ eV. (b) The purple plot corresponds to $p_x=10^{-1}$ eV and the orange one is for $p_x=10^{-2}$ eV.}  \label{Numeric-Analytic}
	\end{figure*}

	In order to check the enhancement effect of different parts of the integrand, we evaluate the integration of Eq.~\eqref{coupling} at the vicinity of the propagator poles
	\begin{eqnarray}
		\label{integralkx}
		\frac{\mathcal{F}^\textbf{p}}{G_a \omega_\textbf{p}}\Big\vert_{\rm from\;the\;propagator\;poles}&=&\int_{-\infty}^{\infty} dk_x\frac{\mathcal{P}(k_x,p_x)}{\tilde{p}_x^2-k_x^2+i\gamma_B}\nonumber\\
		&=& \textrm{not enhanced}
		+\frac{\mathcal{P}(-\tilde{p}_x,p_x)}{2\tilde{p}_x}\int_{-\tilde{p}_x-\epsilon}^{-\tilde{p}_x+\epsilon}
		\frac{dk_x}{\tilde{p}_x+k_x-i\gamma_B/(2\tilde{p}_x)}+\frac{\mathcal{P}(\tilde{p}_x,p_x)}{2\tilde{p}_x}\int_{\tilde{p}_x-\epsilon}^{\tilde{p}_x+\epsilon}
		\frac{dk_x}{\tilde{p}_x-k_x+i\gamma_B/(2\tilde{p}_x)}
		\nonumber \\ &=& 
		\textrm{not enhanced}
		-\frac{2i\epsilon}{\gamma_B}\left[\mathcal{P}(\tilde{p}_x,p_x)-\mathcal{P}(-\tilde{p}_x,p_x)
		\right]\nonumber\\&=&\text{not\;enhanced}~,
	\end{eqnarray}
	in which $\gamma_B=\omega_\textbf{p}\Gamma_{B}$, $\tilde{p}_x^2=p_x^2-m^2$, and ``not enhanced'' implies that the propagator poles don't produce any enhancement on that range of integration. The profile function $\mathcal{P}(k_x,p_x)$ is an even function thus there is no enhancement due to the propagator poles. Considering $\ell\ll L_x$, one can approximate the sinc functions of the profile as Dirac delta functions using $~\text{sinc}\left(\frac{L_x}{2}(k_x+p_x+\frac{1}{\ell})\right)\rightarrow\frac{2\pi}{L_x}\delta(k_x+p_x+\frac{1}{\ell})$ and analytically evaluate the integral as
	\begin{eqnarray}\label{analytic}
		\frac{\mathcal{F}^\mathbf{p}}{G_a\omega_{\textbf{p}}}=\frac{\pi L_x}{4}\left(\frac{1}{-\frac{2p_x}{\ell}-\frac{1}{\ell^2}-m^2+i\gamma_B}+\frac{1}{\frac{2p_x}{\ell}-\frac{1}{\ell^2}-m^2+i\gamma_B}\right)~.
	\end{eqnarray}
	Fig~\eqref{Numeric-Analytic} shows the adaptation between the numeric and analytic calculation of the integration in Eq.~\eqref{Fr} for the scenario which is mentioned in the text; a 1 m cavity with a finesse of $F=10^5$ is exposed to a magnetic field with $\ell=L_x/10\pi$ (the cavity length is 5 times of the magnetic field wavelength) and the coupling constant for the plots of Fig~\eqref{Numeric-Analytic} is $g_{a\gamma\gamma}=10^{-10}\;\text{GeV}^{-1}$. 
	
\end{document}